**Extending Quantum Links: Modules for Fiber- and Memory-Based Quantum Repeaters**


*Peter van Loock\*, Wolfgang Alt, Christoph Becher, Oliver Benson, Holger Boche, Christian Deppe, Jürgen Eschner, Sven Höfling, Dieter Meschede\*, Peter Michler, Frank Schmidt, Harald Weinfurter*

Peter van Loock, Frank Schmidt,
Institute of Physics, Johannes Gutenberg University Mainz, Staudingerweg 7, 55128 Mainz, Germany,

Wolfgang Alt, Dieter Meschede,
Institute of Applied Physics, University of Bonn, Wegelerstraße 8, 53115 Bonn, Germany,

Christoph Becher, Jürgen Eschner,
Fachrichtung Physik, Universität des Saarlandes, Campus E2.6, 66123 Saarbrücken, Germany,

Oliver Benson,
Institut für Physik, Humboldt-Universität zu Berlin, Newtonstr. 15, 12489 Berlin, Germany,
IRIS Adlershof, Humboldt-Universität zu Berlin, Zum Großen Windkanal 6, 12489 Berlin, Germany,

Holger Boche,
Lehrstuhl für Theoretische Informationstechnik, Technische Universität München, 80290 München, Germany,
Munich Center for Quantum Science and Technology (MCQST), 80799 München, Germany,

Christian Deppe,
Lehrstuhl für Nachrichtentechnik, Technische Universität München, 80290 München, Germany,

Sven Höfling,
Technische Physik, Physikalisches Institut und Wilhelm Conrad Röntgen Center for Complex Material Systems, Universität Würzburg, Am Hubland, 97074 Würzburg, Germany,

Peter Michler,
Institut für Halbleiteroptik und Funktionelle Grenzflächen (IHFG), Center for Integrated Quantum Science and Technology (IQ$^{ST}$) and SCoPE, University of Stuttgart, Allmandring 3, 70569 Stuttgart, Germany,

Harald Weinfurter,
Fakultät für Physik, Ludwig-Maximilians-Universität München, Schellingstr. 4, 80799 München, Germany,
Munich Center for Quantum Science and Technology (MCQST), 80799 München, Germany,
Max-Planck Institut für Quantenoptik, Hans-Kopfermann-Str. 1, 85748 Garching, Germany

E-mail: loock@uni-mainz.de, meschede@uni-bonn.de








We analyze elementary building blocks for quantum repeaters based on fiber channels and memory stations. Implementations are considered for three different physical platforms, for which suitable components are available: quantum dots, trapped atoms and ions, and color centers in diamond. We evaluate and compare the performances of basic quantum repeater links for these platforms both for present-day, state-of-the-art experimental parameters as well as for parameters that could in principle be reached in the future. The ultimate goal is to experimentally explore regimes at intermediate distances – up to a few 100 km – in which the repeater-assisted secret key transmission rates exceed the maximal rate achievable via direct transmission. We consider two different protocols, one of which is better adapted to the higher source clock rate and lower memory coherence time of the quantum dot platform, while the other circumvents the need of writing photonic quantum states into the memories in a heralded, non-destructive fashion. The elementary building blocks and protocols can be connected in a modular form to construct a quantum repeater system that is potentially scalable to large distances.





# 1. Introduction

Quantum key distribution (QKD) and related schemes are offering a paradigm change in establishing secure communication: algorithmic security is replaced by physically secure generation of encryption keys.[1] The symmetric keys created by QKD can be used to securely transmit messages between two stations (Alice and Bob) via public channels. Security is warranted by physically detecting any eavesdropping attack. To generate a key, the iconic BB84 protocol [2] employs non-orthogonal quantum states of photons carrying qubit information, while other schemes make use of measuring entangled photon pairs, such as the Ekert protocol.[3] More generally, establishing entanglement of distant quantum objects provides a critical resource for efficient distribution of quantum information, both at short and long distances; applications beyond quantum cryptography, such as distributed quantum information processing and future quantum networks,[4] will also depend on this resource.

Networks based on individual point-to-point links (PPLs) over 50-80 km length have been realized at the metropolitan area level, and even a long distance connecting Beijing and Shanghai (~2.000 km) has been bridged via 32 intermediate stations.[5] So far, however, such networks rely on independent quantum PPLs chained together by "trusted nodes", connecting the links by classical operations ("receive and resend") and thus providing full access to the transmitted bits at each node. Truly long-range quantum links have been realized via satellite channels,[6] yet up to now also the satellites serve as trusted nodes in such schemes. Moreover, since these links require large-scale send-and-receive facilities, it is likely that they need to be combined with "local-area" ground-based quantum networks (of a smaller, intermediate range) as obtainable from the elementary fiber-based schemes presented and discussed here.





At present the main obstacle in establishing large-scale quantum networks are inherent losses of the transmission channels. The current record for terrestrial, fiber-based point-to-point QKD lies in the range of about 400 km.[7,8] As a consequence,[1] secret key rates obtained via direct transmission (without intermediate stations) through an optical quantum channel of length $L$ are effectively limited by the channel transmission efficiency $\eta = \exp(-L/L_{att})$ for large $L$ where $L_{att}$ is the attenuation length of the channel.[9] More precisely, this limit corresponds to a secret key capacity of 1.44 $\eta$ (per channel use and per mode, in units of secret bits[2]).[10] In particular, optical fiber systems feature a loss rate of about 0.2 dB/km (corresponding to $L_{att} = 22$km), limiting useful distances to a few hundred km **(Figure 1)**.

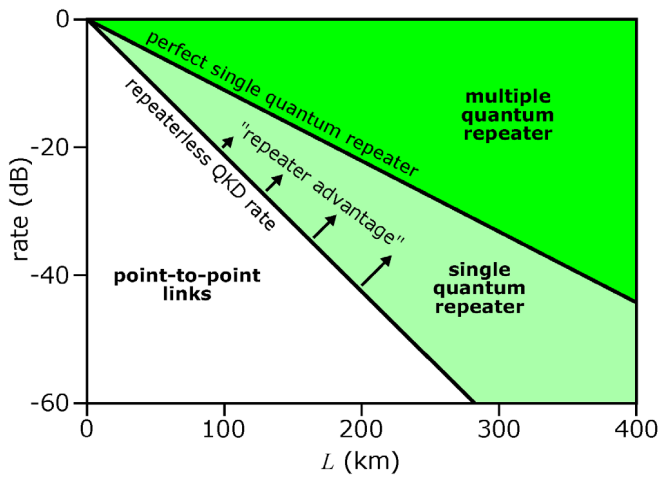

**Fig. 1:** QKD rate in dB (normalized to the protocol's clock rate) as a function of distance in km. Point-to-point protocols scale as $\sim \eta = exp\left(-\frac{L}{L_{att}}\right)$, limited by the "repeaterless" bound.[10] For telecom fibers: $L_{att}$=22km. An ideal "single" quantum repeater with only one middle station scales as $\sim \sqrt{\eta} = exp\left(-\frac{L/2}{L_{att}}\right)$. "Multiple" repeaters may further reduce the effective loss and extend the transmission distance. The exact "repeaterless" bound (secret key capacity) is $-log_2(1-\eta) \approx 1.44\,\eta$ in units of secret bits,[10] where the approximation only holds for sufficiently small $\eta$ (large distances).

There are interesting methods to overcome this limitation without the use of quantum memories by sending fairly simple quantum states (in the form of single photons or optical coherent states) to a detector station placed in the middle of the channel.[11,12] Especially the "twin-field QKD" concept [12] is appealing, as it needs³ neither multiple parallel channel transmissions nor non-destructive measurements with feedforward and multiplexing,[11] but instead only transmission of phase-sensitive single-mode quantum states and their interference at the middle station. Experimental proof-of-principle demonstrations of the twin-field concept were reported very recently.[14,15,16] Both approaches [11,12] reduce the effective channel length by a factor of two, corresponding to an enhanced transmission efficiency of $\sqrt{\eta} = \exp[-(L/2)/L_{\text{att}}]$. However, neither of them has been shown to be scalable to larger distances by further improving the effective transmission. In principle, there are other, all-optical approaches for long-distance, even scalable quantum communication with no need for storing qubits in matter-based memories, but such schemes depend on the engineering of complex multi-photon (entangled) quantum states and a sufficiently close spacing of stations along the channel (every 1-5 km) in order to exploit the sophisticated concept of quantum error correction codes.[17]

---

[1] In combination with transmission losses another limiting factor are dark counts of the detectors. At a distance of 400 km, only ~10 photonic qubits would be transmitted per second when sent at GHz clock rate. Thus, beyond 400 km the optical signals will eventually vanish under dark count noise. In this work, the maximal total distance considered is 400 km, which in the repeater scenario is divided at least into two segments of maximally 200 km length for each.

[2] The factor 1.44 stems from the change of base of the logarithm in the Taylor expansion of the secret key capacity, $-\log_2(1-\eta) = -\frac{\ln(1-\eta)}{\ln 2} = \frac{\eta}{\ln 2} + O(\eta^2)$, where $\frac{1}{\ln 2} = 1.442695$ ... and $\eta \ll 1$.

Therefore, it is currently assumed that the most feasible and promising route towards long-distance quantum communication, while entirely avoiding trusted node configurations, is based upon the use of quantum repeaters [18] that include intermediate stations (typically every 10-100 km) equipped with quantum memories realized by atomic or solid-state qubits. Here, we consider elementary fiber- and memory-based schemes, which we refer to as quantum repeater cells. By storing quantum states for sufficiently long, these schemes allow to enter the rate regime between $\eta$ and $\sqrt{\eta}$ and may serve as modular building blocks for bridging larger distances. Thus, ultimately, true quantum networks based on quantum repeaters should not only eliminate the need to trust the stations along the channels of the network, but also lead to a superior QKD rate scaling with distance when compared with untrusted quantum relays (where each node only measures optical quantum states without storing them). Compared to quantum PPLs chained together by trusted nodes and other forms of quantum relays, genuine repeater-based quantum networks would thus represent a leap both conceptually and quantitatively.

---

³ For a small-scale experiment along the lines of Ref. [11], but circumventing such complications, see Ref. [13].

The first quantum repeater (QR) concepts were proposed already 20 years ago [18] to overcome the distance limitation by distributing, enhancing, and connecting short-range entanglement through local quantum operations and classical communication. In the simplest case, quantum correlations from two entangled point-to-point segments AA' and B'B are connected via a collective Bell-state measurement (BM) at the central "repeater" node A'B', resulting in so-called entanglement swapping to nodes A and B **(Figure 2)**. These larger segments can then be concatenated further in the same way, while a simple multiplication of the channel transmission efficiencies per segment and a propagation and accumulation of errors can be prevented by storing quantum information in quantum memories and applying entanglement purification on many entangled pairs in each segment [18] or incorporating quantum error correction codes into the memory qubits.[17] Overcoming the distance and rate limitations in a scalable fashion, QRs offer highly attractive functionality for future long-range quantum networks.[4]

**Fig. 2:** *Generic QR link for increasing the communication distance. Initially, for each segment AA' and B'B, quantum memories (full circles) are entangled with each other (double red line) over a distance L/2. Via a Bell-state measurement (black box) on the two memories in the central repeater node, the entanglement is swapped to the outer memories A and B separated by distance L. Thus, a new, longer segment is created that is usable for further extensions of the quantum link by repeated concatenation of this procedure including some form of quantum error detection or correction.*

---

[4] For a summary of our graphical symbols to represent QR elements, see Sec. 1 in the Supporting Information.





Experimentally, QRs have remained an enormous challenge up to now.[17,19] A QR constitutes a system based on several different hardware components. Although all necessary components have been demonstrated to some extent individually, combining these into a fully operational (and hence scalable) repeater system is demanding and first experimental demonstrations in this direction are now only beginning to be reported.[20]

One of the most critical hardware components are the quantum memories required to effectively synchronize the arrival of quantum information for further processing at the individual nodes. Depending on the range and the application of the repeater system, the required memory coherence times vary. For example, in order to establish entanglement over 1000 km via a standard QR [18] at least millisecond storage times are needed only to be able to cover the waiting time for a classical signal sent over the total distance. In a fully nested quantum repeater with probabilistic entanglement purification and swapping steps including two-way classical communication, even longer storage times will be required. Deterministic entanglement swapping and quantum error correction of local gate and memory errors may reduce these requirements [17], but most memory systems are still not sufficiently long-lived or fault-tolerant.[21]

Here we analyze small-scale, functional QR systems that may serve as elementary building blocks for experimental QR realizations on a larger scale. Implementations are considered for three different physical platforms, for which suitable components are available: quantum dots, trapped atoms and ions, and color centers in diamond. The aim of these elementary schemes is to experimentally approach a regime at intermediate distances (up to several 100 km) in which the qubit transmission and secret key rates exceed the limits of direct transmission. Based on a simple model we compare the properties of the different platforms capturing the influence of source and memory efficiencies on the repeater performance for each system.





In order to assess and compare the specific capabilities of each platform, we primarily consider the most dominating and distinct effects in a typical elementary QR, namely transmission loss in the fiber channel and memory dephasing at the repeater stations. In addition, we do include source and detector efficiencies, but we omit, for example, detector dark counts. These have a significant impact on secret key rates for larger distances.[5] The overall performance of the source includes an experimentally determined efficiency and a clock (repetition) rate whose influence on the repeater rates depends on the repeater protocol.

The memory quality is given by an experimentally determined coherence time, but the impact of memory dephasing errors on the entanglement fidelity and thus the secret key fraction can be controlled by a freely chosen, so-called memory cutoff time.[22] This means a quantum state is never kept in the memory for longer than a maximal storage time in order to optimize the secret key rates or almost entirely suppress dephasing errors. In our model, for comparison with the dimensionless "repeaterless" bound (secret key capacity), the finally considered secret key rates per channel use and per mode are also dimensionless and not expressed in Hz. Thus, clock rates given in Hz only have an indirect effect on the QR performance via the accumulated dephasing times and the corresponding variations of the required cutoff. We consider two different protocols, one of which is better adapted to the higher source clock rate and lower memory coherence time of the quantum dot platform. The other protocol, however, circumvents the need of writing the transmitted optical quantum states into the memories in a heralded, non-destructive fashion. It will become apparent that for both protocols, in principle, the elementary building blocks can be connected in a modular fashion to construct a QR system that is potentially scalable to larger distances. Let us now first introduce a minimal set of experimental parameters that can be used to quantitatively assess the performance of a memory-based QR system.

---

[5] However, thanks to recent technological developments typical dark count rates can be reduced dramatically (below 1 dark count per second).

## 2. Minimal set of experimental parameters characterizing QR performance

We assess the performance of a single QR cell (as it will be defined in Sec.3) or, similarly, a two-segment QR in a simplified model applicable to all three physical platforms. For this purpose, we choose three experimental parameters that are primarily related to the sources', the detectors', and the memories' efficiencies: the zero-length channel or link coupling efficiency, $P_{link}$, the source/memory clock time $\tau_{clock}$ (time span between two trigger/excitation events or memory write-in and reset time), and the memory coherence time $\tau_{coh}$. The link coupling efficiency $P_{link}$ incorporates the photon creation efficiency, fiber channel in- and out-coupling efficiencies, and, depending on the protocol, a detector efficiency or a memory write-in efficiency; the fiber channel transmission efficiency $\eta$ will be treated separately from $P_{link}$. We consider sources generating true single-photon states as obtainable from initial entangled spin-photon resources. A single photonic qubit that is launched into the fiber channel is encoded into two field modes (typically corresponding to polarization or time-bin encoding). Such single-photon-based two-mode qubits can be easily "rotated" into any qubit state and measured in any qubit basis; for two qubits simple partial Bell-state measurements are available. These single-photon qubit states are also most robust against path length fluctuations along the optical channels and compatible with the stationary matter qubits (as opposed to weak coherent states or other phase-sensitive single-mode states, although also for this case repeater protocols exist [19]). The memory coherence time $\tau_{coh}$ is defined via the time-dependent probability for a random phase flip to occur on a memory qubit, $\frac{1}{2}\left(1 - \exp\left(-\frac{t}{\tau_{coh}}\right)\right)$, see Sec. 2 in the Supporting Information. In addition, we include





a memory cutoff time, i.e. a maximally allowed storage time until any quantum memory is reset and reinitialized. For a summary of the relevant experimental parameters and our notation used throughout the paper, see Sec. 1 in the Supporting Information.

Let us briefly discuss the influence of the finite link coupling and channel transmission efficiencies in an idealized general QR, without errors and for an arbitrary number of stations/segments, on the QR performance, corresponding to a raw rate in the QKD context. We can then compare this with a quantum PPL, i.e., a scheme without the use of quantum memories solely based on direct transmission of quantum states. A single QR segment can be thought of as a quantum PPL over distance $L/n$ when the total channel of length $L$ is divided into $n$ segments. The raw rate in Hz, i.e. the number of quantum bits (secret bits in QKD without errors) per time and per mode, for one segment is then given by

$$\mathcal{R}_{\text{link}}(L/n) = \frac{R_{\text{link}}(L/n)}{N T_0} \, ,$$

where $R_{\text{link}}$ is the overall (dimensionless) link efficiency[6], $T_0$ is the time duration between two channel uses (i.e. the time per use), and $N$ is the number of modes in case that several modes are sent in parallel through the optical channel. In general, $R_{\text{link}}(L/n)$ may exceed unity, but it must necessarily remain smaller than one either for not too short segment lengths (i.e., channel segments with more than 3dB transmission loss for each [10]) in a single-mode link or for an optical encoding based on discrete qubit states, as it applies to our two-mode-qubit-based schemes. This is why we refer to $R_{\text{link}}(L/n)$ as an efficiency and we may decompose it into the two contributions coming from the link coupling and channel transmission efficiencies:

$$R_{\text{link}}(L/n) = P_{link} \, \eta^{1/n},$$

---

[6] Generally, $R_{\text{link}}$ counts the number of raw qubits (secret bits) transmitted per channel use in a multi-mode channel with $N$ modes. It is upper bounded by the multi-mode secret key capacity $-N \log_2(1-\eta) \approx 1.44 \, N \, \eta$.[10]





where, more specifically, the second factor describes the channel transmission in a single repeater segment $\eta^{1/n} = \exp[-(L/n)/L_{att}]$ (i.e., $\eta$ is the probability that a single-photon two-mode qubit remains intact after its parallel transmission over two independent amplitude damping channels of length $L$, while $\sqrt{\eta}$ represents the amplitude damping parameter of a Gaussian single-mode loss channel of length $L$).

If we connect the segments without the use of quantum memories like in a relay, effectively multiplying the efficiencies of the individual segments, we obtain at best $\left(R_{link}(L/n)\right)^n = (P_{link})^n(\eta^{1/n})^n = (P_{link})^n \, \eta$. Since this scales with distance like a PPL over the whole channel, we may just remove the intermediate stations to obtain $R_{link}(L) = P_{link} \, \eta =: R_{PPL}(L)$. This link efficiency for the total two-mode PPL, up to a factor of 1.44 and for small $P_{link} \, \eta$, can also be identified as a "realistic repeaterless" bound for a single-mode channel of length $L$ including a finite link coupling efficiency for the quantum PPL between Alice and Bob with finite source, fiber coupling, and detector efficiencies at Alice's and Bob's stations. For the raw rate in Hz (per mode) obtainable over the whole channel, we can now also write $\mathcal{R}_{PPL}(L) = R_{PPL}(L)/NT_0 = (P_{link} \, \eta)/NT_0$. In this case, if Alice directly sends a qubit to Bob over the entire distance, she will use $N=2$ modes for a two-mode-encoded photonic qubit and she may also send many qubits sequentially at a high source clock rate $(\tau_{clock})^{-1} \sim$ GHz such that the final rate $\mathcal{R}_{PPL}$ is ultimately limited only by $\eta$ since $T_0 = \tau_{clock}$ (also assuming sufficiently fast detectors at Bob's station).

Once quantum memories are employed at the intermediate stations, in principle, a raw rate in Hz (per mode) for the total distance scaling as $\mathcal{R}_{QR} \sim (P_{link} \, \eta^{1/n})/NT_0$ can be approached (at fixed $n$), which corresponds to an expression similar to that for the rate in a single QR segment. The quantity $P_{link}$ is once again the link coupling efficiency related with a single





repeater segment and recall that we do not consider additional success probabilities from entanglement purification and swapping in the present discussion on an idealized QR. However, $P_{link}$ should now also contain any inefficiencies related to the light-matter interface or the memory write-in for one segment. Even more important, compared with a memoryless quantum PPL bridging the total distance, the time unit for one channel use $T_0$ (as only for a PPL uniquely defined and coinciding with the source/detector clock time) will be significantly larger than a source clock time $\tau_{clock}$. For the memory-based QR, depending on the specific protocol, $T_0$ must include the local memory write-in and reset times (~MHz$^{-1}$) and the necessary waiting times for classical signals announcing successful quantum state transmissions. Thus, although typically one also has $N=2$ modes for the optical qubits, beating even the realistic "repeaterless" bound expressed in Hz requires a sufficiently long distance such that the superior scaling of $\eta^{1/n}$ dominates over the inferior "clock rate" of the memory-based repeater. So it is important to recognize that even the ideal memory-based QR, compared to a quantum PPL with fast sources and detectors, starts with a "repeater disadvantage", and only for sufficiently large distances can this be converted into a "repeater advantage". If errors are included, no longer all transmitted qubits (when employed for QKD) can be turned into secret bits. Related with this, for large distances, the QR rates drop further due to the need of probabilistic quantum error detection (such as entanglement purification) on higher repeater levels (alternatively, as said before, quantum error correction may be employed for all local gate and memory errors).

Note that all-optical quantum repeaters (at least those that work entirely without feedforward operations at the intermediate stations) can, in principle, operate at the same clock rate as a direct-transmission PPL. However, not only do we need rather complicated encoded states for this approach, but typically (though not necessarily) many optical modes $N>2$ are required to transmit a logical qubit. Therefore, also in this case, sufficiently many segments have to be





concatenated to benefit from the better effective transmission per segment, $(R_{\text{link}})'(L/n)$, compared to the long-distance PPL that works with $N=2$. Such a better effective transmission due to quantum error correction at every station requires sufficiently short segment lengths, as opposed to the schemes we consider below. For short segment lengths, as already mentioned above, non-qubit-based schemes would in principle even allow for a "link efficiency" greater than one corresponding to the transmission of more than a single qubit (secret bit) per channel use.[7] A unique exception is the twin-field QKD concept, for which we also have a high clock rate, only limited by lasers and detectors, and even just a single mode $N=1$ for the optical transmission. However, this approach is not known to be scalable beyond $\sqrt{\eta}$.[8]

To conclude, beating the (realistic) dimensionless "repeaterless" bound by means of a multi-mode memory-based quantum repeater with an effective overall transmission efficiency $R_{\text{QR}}$, i.e. effectively exceeding the overall efficiency of a multi-mode direct-transmission PPL,

$$R_{\text{QR}}(L) \; > \; 1.44 \, N \, P_{link} \, \eta \; = \; 1.44 \, N \, R_{\text{PPL}}(L) \; \gtrsim \; (N/2) \, R_{\text{PPL}}(L) \; ,$$

is the minimal requirement even for a small-scale quantum repeater module to eventually be able to obtain better rates $\mathcal{R}$ in Hz for large-distance quantum communication with many modules than what is obtainable via a long-distance PPL. Here, $N$ is the number of modes and $R_{\text{PPL}}(L) = P_{link} \, \eta$, as introduced earlier, refers to a two-mode direct-transmission PPL that covers the total channel and employs no quantum memories at all. Thus, here the link coupling efficiency contains only source (with fiber in-coupling) and detector (with fiber out-

---







coupling) efficiencies, $P_{link} = P_{source} \, \eta_{det}$. The factor 1/2 in the lowest bound above has been included to stress that $R_{\mathrm{PPL}}(L)$ represents a two-mode link efficiency. The bound in the middle is the (realistic[9]) multi-mode "repeaterless" bound for large $L$. In other words, overcoming the dimensionless bounds with a small, elementary repeater is the first necessary condition to be met for an experimental demonstration of in-principle scalable quantum repeater functionality. In our schemes, the QR stations are connected by optical two-mode channels, hence *N=2.* In this case, overcoming the direct-transmission efficiency bound expressed by a two-mode PPL corresponds to $R_{\mathrm{QR}}(L) > R_{\mathrm{PPL}}(L) = P_{link} \, \eta$. In our quantitative comparison later (Figs. 5 and 7), we will consider as a figure of merit the secret key rate, SKR, in a memory-based QR scheme per channel use and per mode. Up to the secret key fraction factor that includes the effect of the dephasing errors for a chosen QKD protocol (see Section 2 of the Supporting Information), SKR then corresponds to $R_{\mathrm{QR}}(L)/2$. The relevant benchmarks will be the ideal "repeaterless" bound (single-mode secret key capacity), $-\log_2(1 - \eta)$, and SKR for a "realistic" but error-free PPL (per channel use and per mode), $R_{\mathrm{PPL}}(L)/2 = P_{link} \, \eta/2$. Yet ultimately, a comparison must rely on rates in Hz, per time and per mode: $\mathcal{R}_{\mathrm{QR}}$ versus $\mathcal{R}_{\mathrm{PPL}}$.

To sum up, for a given channel transmission efficiency (with $L_{\mathrm{att}} = 22\mathrm{km}$), we consider three fundamental parameters:

*(1) the link efficiency* $R_{\mathrm{link}}$, which is composed of the link coupling efficiency $P_{link}$ (now also including memory efficiencies) and the channel transmission efficiency per segment $\eta^{1/n}$,

(2) *the memory coherence time* $\tau_{coh}$, which can influence both the repeater raw rates and the secret key fraction in the QKD context, and

---

[9] Note that in our notation for $P_{link}$ we do not make a distinction between links of different mode numbers.





*(3) the clock time $\tau_{clock}$*, which, depending on the protocol, can have a significant impact even on the dimensionless repeater rates, namely indirectly in the presence of memory dephasing.

In the following we will discuss in detail several variants of small-scale proof-of-principle repeater protocols which can be classified into basically two distinct classes: node sends photons ("NSP") and node receives photons ("NRP"). For each protocol we will then specify the particular form of the above three fundamental parameters, especially decomposing the link efficiency into further experimental parameters depending on the protocol. Eventually we will be able to insert particular values for each of the three hardware platforms in order to compare their possible present and future repeater performances.

## 3. QR cell: A generic experimental system showing QR functionality

Before introducing the basic concept of a QR cell in detail, and applying it to two different protocols and three different physical platforms, let us start by summarizing the overall concept for establishing a QR within our framework:

- A quantum channel is realized by an optical fiber.
- Intermediate stations along the channel include sources of single/entangled photons or spin-photon entanglement, beam splitters, detectors, possibly wavelength converters.
- The "repeaterless" bound limits the (secret key) rates in point-to-point communication (direct transmission without intermediate stations).
- The QR segments create entanglement of two spatially separated quantum memories connected by a direct quantum channel.
- The QR cells consist of two half QR segments with a central QR node containing quantum memories.

As described in the introductory part, the focus here is on fiber channels with a fixed channel attenuation. We omit the quantitative effect of wavelength converters which, in our model,





could be absorbed into $P_{link}$ via a wavelength conversion efficiency. While Fig.2 above shows how entangled QR segments, once they are available, can be connected by entanglement swapping to increase the distance of a QR, **Figure 3** illustrates how a single QR segment itself, defined as an entangled pair of quantum memories located at neighboring repeater stations, may be established via an optical BM on two photons (two qubits) emitted by the two quantum memories placed each at the end points.[10]

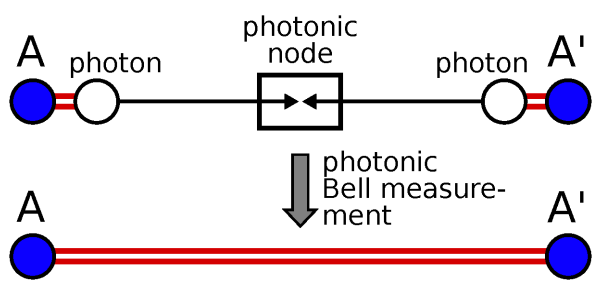

*Fig. 3:* Entanglement creation within a QR segment (with QR nodes sending photons like in the "NSP" protocol below). At the end nodes spin-photon entanglement (full-open pair of circles) is generated. An optical Bell-state measurement on photons arriving at the central photonic node produces entanglement of the end nodes. This configuration does not yet exploit the storage capabilities of the quantum memories, since the photons need to arrive simultaneously at the middle station.

## 3.1. Protocol 1: Node sends photons

### 3.1.1. Model, parameters, modularity, rate analysis

One of the simplest, most generic protocols promising to show the functionality of a memory-based QR system was put forward by Luong et al.[23] This protocol, which we refer to as NSP ("Node Sends Photons") protocol, is based on an arrangement that we will call a "quantum repeater cell" (QR cell). Generally, this is an elementary structure that contains the minimal set of components required to show the functionality of a memory-based QR scheme, thus allowing to analyze schemes that can, in principle, overcome the "repeaterless" bound. An additional important property of a QR cell is that concatenation of QR cells renders the system (if, ideally, only affected by channel loss), in principle, scalable **(Figure 4)**. This extra feature is needed, as we know that the "repeaterless" bound can be overcome in a restricted (not fully scalable) sense via a middle station not equipped with quantum memories.[11,12] The NSP protocol relies on only a few generic parameters, whose impact on the QR performance can be clearly identified. It thus allows to compare different hardware platforms, including a qualitative and quantitative assessment of their relative strengths and weaknesses.

---

[10] For a summary of our graphical symbols to represent QR elements, see Sec. 1 in the Supporting Information.

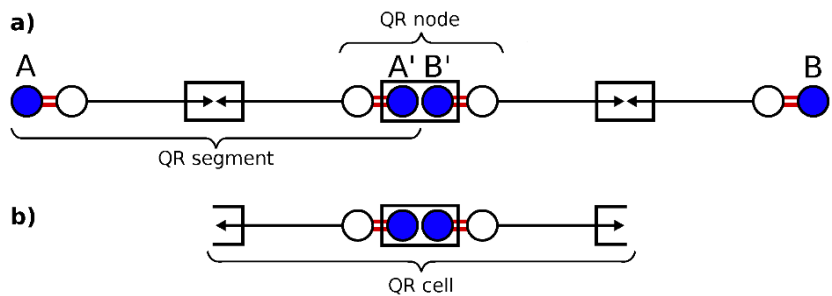

*Fig 4:* (a) Full QR link with two QR segments (NSP) like in Fig. 3. (b) QR cell (NSP) with two half QR segments and a central node for storage as a minimal element for exploiting memory capability. The pair of quantum memories at the central node enables a valid Bell-state measurement also when the left and right half segments become entangled at different times.

For a functioning QR cell (Fig.4b) the central node, equipped with a pair of quantum memories, is crucial. It allows to asynchronously establish effective entanglement in the two half segments, although an entangled state will never be physically shared between the end points of a QR cell. Instead, one would measure the optical signals emitted from the central node at the end points of the cell to establish correlations and obtain a secret key. The specific feature of the NSP protocol for the QR cell is that at the central QR node quantum states with spin-photon entanglement are locally created and then the photons are coupled into the communication channels, i.e. the node sends photons towards the detectors placed on the left and right ends of the cell (Fig.4b). The concatenation of several QR cells then involves two-photon interferences to perform optical two-qubit BMs at the "photonic nodes" (Fig.4a). Note that similar elementary QR schemes with a single QR node emitting and sending photons were considered in Refs. [24,25] (considering a range of experimental parameters similar to Ref. [23], however, including additional memory cutoffs, being adapted to the





specific hardware platform of NV centers, and, in Ref. [25], incorporating the twin-field QKD concept [12] based on single-photon interference).

Let us discuss the underlying model for a QR cell with the NSP protocol in more detail. A single QR cell (Fig.4b) of total length $L$ is composed of a central memory station placed in the middle between two receiving stations each equipped with photon detectors. The conceptually simplest scenario is when the two quantum memories each emit a single photon in two polarization modes entangled with the memory internal state. One photon is sent to the left receiver and the other photon to the right receiver (Fig.4b). The probability for each photon to arrive at its intended detector after travelling over a channel distance $L/2$ is $\exp[-(L/2)/L_{att}] \equiv \sqrt{\eta}$. Without the use of quantum memories both detectors must click simultaneously for the transmission to succeed, which happens with a probability $\sqrt{\eta}^2 = \eta = \exp(-L/L_{att})$ corresponding to the direct-transmission efficiency over a distance $L$. Thus, a single photon could be equivalently sent directly from left to right without the central station. However, by employing quantum memories, once the middle station is informed about the detection of one photon left or right, the respective memory is kept and for the other light-memory pair further attempts are made to eventually have a second photon arriving at its detector and being detected. A final BM on the two quantum memories, effectively swapping the entanglement of the two spin-photon pairs onto the two successfully distributed photons, establishes correlations between the two detectors such that a secret key can be shared provided that non-commuting observables were measured at the photon detectors (like in a BB84 protocol). Thanks to the memories, in principle, the transmission probability for the total distance $L$ then scales as $\sqrt{\eta}$, corresponding to an effective transmission over only half the distance $L/2$.





The most extreme scenario in a QR cell would be to attempt distributing effective entanglement by sequentially (rather than simultaneously) sending photons entangled with memory qubits to the left and to the right (e.g., first to the left), and start sending those photons entangled with a second spin (e.g. the right one) only when the arrival of a photon belonging to the first spin (e.g. arriving at the left detector) was confirmed and the first spin qubit (e.g. the left quantum memory) was determined to be held for storage. Such an approach can be experimentally useful, because the central node may no longer require two distinct memory systems (with the typical example of a single NV center whose nuclear spin with coherence times of the order of seconds allows for efficient storage and whose electron spin with coherence times of the order of milliseconds can be employed as an interface to the optical communication channel [24,25]; another example would be an ion-based quantum memory composed of two ion species where one is adapted for storage and the other for light-matter interfacing [26]).

The effective transmission probability $R_{QR}$ is related to the inverse average number of attempts it takes for successfully transmitting the photons to both ends. However, besides this average number, the ultimate secret key (or qubit) rate of a repeater scheme expressed in secret bits (or qubits) per second, $\mathcal{R}_{QR}$, also depends on the actual duration per attempt (recall the discussion in Sec.2). Moreover, the longer a single attempt takes, the smaller the number of attempts becomes that can be executed well within a given quantum memory's coherence time. In the NSP protocol, the duration per attempt is distance-dependent, because any new attempt can only be initiated when the classical signal from the detector has been received. Thus, the total duration of a single attempt including quantum and classical signal transmissions is $T_0 = \frac{L}{c}$ for the QR cell (Fig.4b) and $T_0 = \frac{L}{2c}$ for the two-segment setup in Fig.4a assuming the same total distance $L$ in either case. Hence a correspondingly larger clock





rate $(\tau_{clock})^{-1}$ for emitting the photons would not be beneficial at all and so this experimental parameter is less relevant for the NSP protocol.

For the QR cell in the NSP protocol (Fig.4b), we have the link coupling efficiency $P_{link} = P_{source}\,\eta_{det}$ where $P_{source}$ includes all efficiencies related to a source emitting photons entangled with a spin memory and coupling them in (and eventually out of) the fiber channel, i.e. it is the probability to get a photon into and out of a single-mode fiber channel per trigger/excitation event, and $\eta_{det}$ is the detector efficiency. Constructing two QR segments like in Fig.4a with the NSP protocol corresponds to $P_{link} = \frac{1}{2}(P_{source})^2\,(\eta_{det})^2$, because one segment is successfully bridged only when both sources at its end points create photons that are both detected at the photonic node in the middle (the factor ½ takes into account the efficiency of a standard partial, beam-splitter-based two-photon two-qubit BM). However, the time duration per attempt for one segment of the two-segment scheme (Fig.4a) is half as big as that for the QR cell (Fig.4b) at any given total distance $L$, as mentioned above.

In addition to the three experimentally determined parameters $P_{link}$, $\tau_{clock}$, and $\tau_{coh}$, we include a memory cutoff parameter imposing the rule that quantum states will never be stored for a longer time than given by the cutoff. [22] In other words, the QR protocol is aborted and started from scratch as soon as a quantum memory's storage time has exceeded the imposed storage limit. The memory cutoff can be freely chosen. Our analysis is based on the experimental parameters for the three platforms as given in the Tables below. **Table 1** refers to the state of the art presenting the currently available, realistic values for each platform. **Table 2** shows potential future parameter values, i.e. an idealization compared to the state of the art. Nonetheless, the latter are physically reasonable and not fundamentally unobtainable.





**Table 1.** Currently available experimental parameters for the three QR platforms: color centers (NV and SiV), quantum dots, ions (Calcium) and atoms (Rubidium).

| Parameters | $P_{link}$ [percent] | $(\tau_{clock})^{-1}$ [MHz] | $\tau_{coh}$ [ms] |
|---|---|---|---|
| Platform | | | |
| NV centers[a] | 5 | 50 | 10 |
| SiV centers[b] | 5 | 30 | 1 |
| Quantum dots[c] | 10 | 1000 | 0.003 |
| Ions[d] (Calcium) | 0.4 | 0.06 | 0.8 |
| Atoms[e] (Rubidium) | 70 | 5 | 100 |

[a] Refs. [24,25], [b] Refs. [27,28], [c] Refs. [29,30,31], [d] Ref. [32], [e] Refs. [33,34]

**Table 2.** Potentially available future experimental parameters for the three QR platforms: color centers (NV and SiV), quantum dots, ions (Calcium) and atoms (Rubidium).

| Parameters | $P_{link}$ [percent] | $(\tau_{clock})^{-1}$ [MHz] | $\tau_{coh}$ [ms] |
|---|---|---|---|
| Platform | | | |
| NV centers[a] | 50 | 250 | 10000 |
| SiV centers[b] | 50 | 500 | 100 |
| Quantum dots[c] | 60 | 1000 | 0.3 |
| Ions[d] (Calcium) | 10 | 1 | 1 |
| Atoms[e] (Rubidium) | 70 | 100 | 1000 |

[a] Refs. [24,25], [b] Ref. [20], [c] Refs. [31,35], [d] Ref. [36,37], [e] Refs. [33,34]

The future parameters of NV centers are obtained by extrapolating the values of Refs. [24,25] for the link coupling efficiency and clock time, and assuming a $^{13}$C nuclear spin for the memory. Similar assumptions are made for the SiV centers based on Refs. [20,27,28].





Compared to NV centers, the SiV platform has the advantage of not only allowing for efficient quantum storage via the nuclear spins but also providing a potentially more efficient photon-spin interface (with higher cooperativities available); though a drawback of SiV is the need for very low temperatures (below 500 mK).[38]

For the quantum dot platform, based on experimentally achieved quantum dot photon-collection efficiencies of 60% [30] connected with a near Gaussian beam profile which is preferential for large fiber in-coupling efficiencies, we estimate the link coupling efficiency $P_{link}$ to 10% (Table 1). Anticipating improvements in photon collection efficiencies up to 90% together with improved fiber-coupling efficiencies, we assume that a possible future value of $P_{link}$ is 60% (Table 2). Regarding the clock times, we estimate spin-preparation times in a quantum dot to be in the few 100 ps regime, and together with reported radiative recombination times also in the range of a few 100 ps,[31] we expect achievable clock rates of 1000 MHz for a quantum-dot-based non-classical light source.

We assumed fairly good experimental parameters for the Rubidium-atom platform compared with those assumed for Calcium ions. The presently available values for $P_{link}$ and $\tau_{coh}$ refer to current experiments with Rubidium atoms in a cavity.[33,34] More specifically, atomic eigenstates can be chosen for the qubit encoding such that the effect of external magnetic fields is significantly reduced. This way coherence times above 100 ms have been measured.[33]

The performance of a QR may be quantified in a meaningful way by the secret key rate that can be obtained for a given length $L$ of the quantum channel connecting the two parties Alice and Bob. The advantage of using the secret key rate as a figure of merit is that it incorporates both the efficiency and the quality (or fidelity) of the quantum state transmission at the same





time. A high efficiency, i.e., a high (effective) transmission probability or raw rate leads to an increasing secret key rate, whereas a low fidelity, i.e. a high error rate, results in a decreasing secret key rate (typically incorporated via a "secret key fraction"). In our rate analysis, we shall consider, on the one hand, secret key rates in an entanglement-based BB84-type scheme, for which optimal memory cutoffs exist, since a cutoff chosen too small will reduce the raw rate and a cutoff chosen too large will lead to a stronger accumulation of dephasing errors reducing the secret key fraction. In other words, the infidelities from the finite coherence times of the memories, eventually becoming manifest as an infidelity of the effective entangled state shared between Alice and Bob after the BM on the memory qubits, are mapped onto a reduced secret key fraction for a BB84 QKD scheme (see Sec. 2 of the Supporting Information).

On the other hand, in an alternative picture independent of QKD, we shall only consider the raw rate (without inclusion of dephasing errors) by choosing the cutoff sufficiently small in order to almost entirely suppress dephasing errors and keep the final fidelities of the (effective) entangled state above a certain value such as 0.95. This means the maximally allowed storage time is chosen well below the memory's coherence time for the loaded memory at the central station waiting for the second transmission to succeed. More details can be found in Sec. 3 of the Supporting Information.

It should be stressed that our simplified models do not entirely capture intrinsic effects arising from specific memory errors (beyond pure dephasing) and other error sources for a given hardware platform, such as an imperfect initial spin-photon state prior to its storage-time-dependent dephasing, imperfections of the final two-spin two-qubit BMs, and detector dark counts. All these additional error sources lead to effective entangled states that are random mixtures of four instead of just two Bell states (see Sec.2 of the Supporting Information)





resulting in secret key rates eventually dropping to zero beyond certain distances. An advantage of the models, however, is that we are able to use only very few simple parameters to compare QR schemes employing different hardware realizations with different error mechanisms for the preparation and storage of quantum states. We can then clearly identify which parameter influences the (still to some extent idealized) QR performance in a certain way, mainly manifesting itself in the rate-vs-distance plot of Fig.1 as a negative offset, i.e. a down shift of the curve due to link coupling inefficiencies, and an increased slope, i.e. an additional distance-dependent rate reduction due to memory inefficiencies.

### 3.1.2. Results and comparison for different platforms

The resulting raw and secret key rates calculated for our model in the case of the NSP-QR cell (as illustrated by Fig. 4b) with the different hardware platforms can be seen in **Figure 5**. The upper part shows the raw rates RR for distributing effective entangled states with a fidelity of at least 0.95 for current (left) and future (right) experimental parameters. The lower part shows the corresponding secret key rates SKR. All rates (in dB) are per channel use and per mode (recall the discussion at the end of Sec. 2).[11]

With current parameters, only the Rubidium-atom platform enters the repeater regimes. For future values, as calculated, both the platform based on Rubidium atoms and that based on color centers enter the repeater regimes at about 100 km and exhibit a slope increase, i.e. a more rapid decline of the rate, starting at around 200 km for NV centers and Rubidium atoms (the decline for Rubidium is faster here because of the ten times smaller memory coherence time, see Table

---

[11] The apparent discontinuities in the RR curves occur, because the cutoff parameter must always be readjusted depending on distance in order to ensure that a fidelity of at least 0.95 is attained (in particular, the discontinuities are not the result of a numerical simulation; our rate calculations are entirely analytical). For calculating SKR always a fixed cutoff parameter was chosen, although there are actually different optimal cutoffs for different distances. The fixed cutoff was chosen such that over the entire regime of distances, rates cannot be much further improved through cutoff variations.





2). The slope increase for SiV centers occurs at even smaller distances due to a memory coherence time assumed to be smaller by another factor of ten. Apparently, the slope of the rates is clearly connected to the memory efficiencies. The plots cover distances up to 400 km and the curves may be extrapolated to larger distances. However, recall that detector dark counts and some other imperfections that could make the rates eventually drop to zero are not included here. The negative offset from the "repeaterless" bounds at zero distance is related to the link coupling efficiency which, for example, is assumed to be worst for the future case of Calcium ions (see Table 2). The platforms based on Calcium ions and quantum dots, as calculated here for the NSP protocol, do not enter the repeater regimes at all, not even for future parameters and not even with regards to the realistic "repeaterless" bounds as a benchmark. Some curves drop faster than the "repeaterless" bound, which seems contradictory. However, note that even when the very first qubit distribution attempt is successful both memories are already subject to dephasing for one time unit. For platforms with insufficient coherence times, this results in an even steeper decline of the secret key rates compared to the "repeaterless" bound, although the $\eta$-scaling could be formally attained via the raw rate by not storing the quantum states at all, i.e., setting the cutoff value to zero (see Supporting Information). All this will become different for another protocol below (NRP) for which, in particular, all platforms are able to access the repeater regimes.





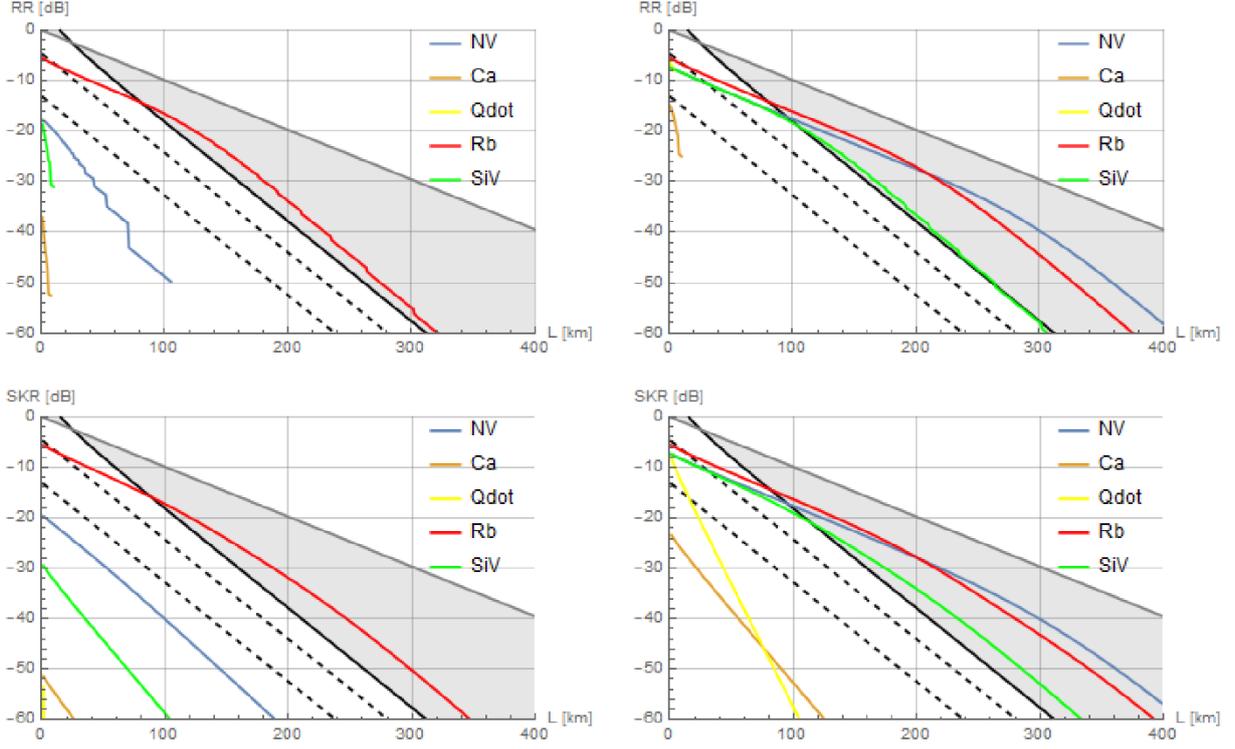

**Fig. 5:** *Secret Key Rates (SKR) and High-Fidelity Raw Rates (RR) for a small NSP-based QR scheme (QR cell). The bottom plots show SKR in dB as a function of the total distance L in km for experimental parameters as currently available (left) and as potentially available in the future (right). The top plots show RR in schemes where the entangled states effectively created over the total distance L have a fidelity of at least 0.95 (left: current parameters, right: future parameters). Curves that are disappearing beyond certain distances (or completely missing for quantum dots) no longer (never) exceed F=0.95. The different platforms correspond to NV (violet) and SiV (green) centers, Calcium ions (brown), Rubidium atoms (red), and quantum dots (yellow). The light grey area illustrates the (secret key) rate regime between ~η (curve in bold black: "repeaterless" bound) and $\sqrt{\eta}$ (line in dark grey: optimal rate for QR cells or two-segment QR schemes). The bold black dashed lines represent the realistic "repeaterless" bound $P_{link}\eta/2$ (direct transmission via PPL) with finite link efficiencies $P_{link} = 0.1,\ 0.7$.*

For the NSP protocol, besides a single QR cell (Fig. 4b), there is also the variant of a QR with two full segments (Fig. 4a). As discussed before, for equal total distance *L,* the two-segment scheme has a smaller elementary time unit compared to the QR cell ($T_0 = \frac{L}{2c}$ versus $T_0 = \frac{L}{c}$). However, at the same time, the two-segment scheme has a smaller link coupling efficiency ($P_{link} = \frac{1}{2}(P_{source})^2\ (\eta_{det})^2$ versus $P_{link} = P_{source}\ \eta_{det}$). For comparison and completeness, we present the rates of the two-segment scheme in Sec. 4 of the Supporting Information. One can see that it performs slightly worse compared to the QR cell. In all plots the secret key rates can sometimes be greater than the raw rates, which again seems contradictory. However, note that for the secret key rates, the optimized memory cutoff (which must neither be too small nor





too large to prevent a too small raw rate or a too small secret key fraction, respectively) typically leads to a worst-case fidelity much lower than the minimal fidelity of 0.95 allowed for the calculation of the raw rates alone (requiring a very small memory cutoff to almost entirely suppress dephasing errors).

**3.2. Protocol 2: Node receives photons**

3.2.1. Model, parameters, modularity, rate analysis

In order to potentially benefit from a higher source repetition rate as available from the quantum dot platform, we shall consider an alternative NRP ("Node Receives Photons") protocol **(Figure 6)**. In this protocol, photons are sent from two sending stations to the central memory station where the arrival of a photonic qubit is non-destructively (e.g. by a linear-optics photonic BM teleporting the arriving photonic qubit to the memory qubit) detected before or while it is "written into" the memory. At any failure event, the next photon pulse can be processed with a delay only depending on the repetition rate of the source or depending on the typically longer write-in and reset times of the memory. In this case, the duration per attempt $T_0$ corresponds to the clock time of the source or the write-in time (which would be the same if spin-photon entanglement is employed both for preparing BB84-encoded photons at the source and for teleporting them into the memories) and is independent of the channel distance.





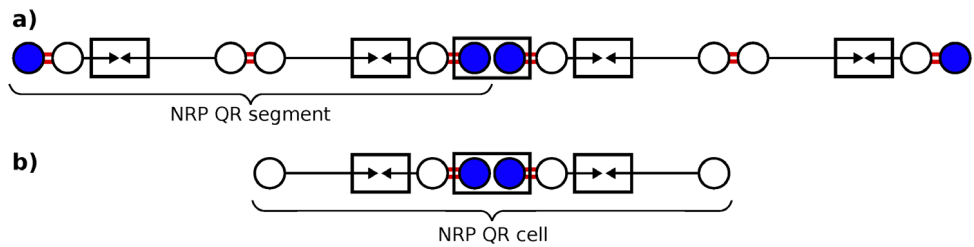

**Fig 6:** *(a) Full QR link with two QR segments incorporating the NRP concept. The BMs in Fig.4a are now replaced by Bell-state sources. (b) QR cell consisting of two half QR segments and a central node for storage as a minimal element for exploiting memory capability. As opposed to the QR cell in Fig.4b, here the quantum memories "receive" photons from two sending stations; whether a photon has arrived must be confirmed by a non-destructive measurement on the qubit, here realized by a photonic BM on a "local" photon emitted from the memory (open circle) and the photon transmitted through the channel. As before, the final BM on the memories can be valid also when the QR segments become entangled at different times.*

A QR cell now still has a central node equipped with quantum memories, but at the end points there are no longer detectors, but sources for optical quantum states such as BB84-encoded single-photon-based qubits (Fig.6b). The memory node now receives the photons which may be realized by a direct and heralded write-in mechanism (such as those of Refs.[39,40,41]), for which certain write-in inefficiencies and infidelities would apply, or by first preparing spin-photon entangled states at the central node and then coupling the photons near the memories locally with the arriving photons coming from the left and right sources (by an optical BM, see Fig.6b). Similar to the NSP protocol, also QR cells based upon the NRP protocol can be concatenated in order to scale up the QR system to larger distances (Fig.6a). The "photonic nodes" where the half segments meet are now no longer performing BMs like in the NSP case, but are instead equipped with entangled photon pair sources (Fig.6a). Compared to the NRP-based QR cell here, a similar elementary QR scheme with a single QR node receiving photons, for BB84-encoded photonic qubits equivalent to what is referred to as measurement-device-independent QKD [42,43] assisted by a quantum-memory-based middle station, was considered in Refs.[44,45,46,47] (again mainly adapted to the specific hardware platform of NV centers, but also presenting comparisons with other platforms in Ref. [46] and





incorporating the idea of a deterministic final BM on the electronic and nuclear spins of a single NV center in Ref. [47]).

In order to keep memory dephasing errors small and the fidelity of the effective entanglement shared between Alice and Bob above a certain minimum, in the NSP protocol, for an increasing $L$ a decreasing number of attempts can be executed at a given memory coherence time because of the $L$-dependence of a single attempt's duration and the growing storage time needed per transmission attempt. In the NRP-protocol-based QR cell (Fig.6b) this $L$-dependence disappears, since the quantum signals are sent to, and no longer emitted from, the quantum memories. The memory cutoff can be chosen independent of distance and the time duration per transmission attempt can be made arbitrarily small by increasing the repetition rate of the sources up to the local memory write-in and reset times. This means the cutoff (expressed by the number of allowed attempts during one storage cycle) can be chosen much higher resulting in larger raw rates. Moreover, this way the memories have less time to be subject to dephasing during a given number of attempts leading to a larger secret key fraction. Generally, the NSP and NRP protocols have both their benefits and disadvantages. The NSP protocol does not require a non-destructive detection of an arriving photon or an efficient heralded write-in mechanism, but the memory station has to wait for the classical signals from the receiving detector stations. In contrast, the NRP protocol relies on a non-destructive measurement or any other means to non-destructively write the incoming "flying qubit" into a "stationary qubit" in a heralded fashion; however, there are no extra waiting times for classical signals (as long as we consider the elementary QR cell of Fig.6b). In addition, the NRP scheme inherits all benefits of "measurement-device-independent" QKD with an untrusted middle station receiving and measuring the quantum states coming from two outer sending stations.[42,43,44,45,46,47] For the rate analysis of the NRP-based schemes, the main experimental parameters taken into account are the same as for the NSP-based cases, i.e., a





link coupling efficiency $P_{link}$ and a memory coherence time $\tau_{coh}$, while the source/memory

clock time $\tau_{clock}$ may have an actual impact only now for the NRP case.

For the QR cell in the NRP protocol (Fig.6b), we now have $P_{link} = P_{source}\,P_{write}$ where

$P_{source}$ again includes all efficiencies related to a source emitting photons (this time prepared

in BB84-states) and coupling them into (and eventually out of) the fiber channel. The

parameter $P_{write}$ represents the probability for successfully writing a photonic qubit arriving

at the central node into the respective memory. If a spin-photon entangled state and a linear-

optics BM are exploited for this in order to teleport the arriving photonic qubit to the memory

spin qubit (see Fig.6b), we have $P_{write} = \frac{1}{2}\,P_{source}\,(\eta_{det})^2$ where $P_{source}$ specifically refers

to the generation of a spin-photon entangled state. Note that if the BB84-encoded photons

were produced in a similar fashion (via initial spin-photon entanglement) with the same

source efficiency $P_{source}$, we would obtain the link coupling efficiency $P_{link} =$

$P_{source}\,P_{write} = \frac{1}{2}(P_{source})^2\,(\eta_{det})^2$, which actually coincides with that of the NSP-based

two-segment QR (Fig.4a), because in terms of the link couplings the two schemes become

identical when the photonic nodes in the middle of each segment of the NSP scheme both

move to the central node right next to the memories (except that the "local" photons may no

longer require fiber coupling). For other write-in methods [39,40,41] we may just directly insert

numbers for $P_{write}$. Although the two-segment concatenation of NRP-based QR cells and half

segments (Fig.6a) demonstrates that the basic modules can be systematically combined to

build an in-principle scalable QR system, we shall not consider this scheme in our rate

analysis. As opposed to the QR cell in Fig.6b, the combined scheme in Fig.6a does require

classical communication to inform the two central memories about the successful loading of

their memory counterparts with photons originating from the same entangled photon pair, and





thus it will have smaller rates than the QR cell alone (however, see Ref. [48]). More theoretical details can be found in Secs. 2 and 3 of the Supporting Information.

### 3.2.2. Results and comparison for different platforms

The resulting raw and secret key rates calculated for our model in the case of the NRP-QR cell (as illustrated by Fig. 6b) with the different hardware platforms can be seen in **Figure 7**. The upper part again shows the raw rates for distributing effective entangled states with a fidelity of at least 0.95 for current (left) and future (right) experimental parameters. The lower part again shows the corresponding secret key rates. All rates (in dB) are again per channel use and per mode (recall the discussion at the end of Sec. 2)

This time we observe that with future parameters all platforms enter the repeater regime for the secret key rate. Moreover, for the simple model used in the rate calculations (no dark counts and no depolarizing errors), all platforms except the Calcium ions achieve a rate slope $\sim\sqrt{\eta}$ over the entire distance of 400 km as shown, thus fully exhibiting the repeater advantage. This holds in particular for the quantum dot platform that, though having the worst memory coherence time, can fully benefit in the NRP protocol from the highest clock rate (see Table 2). With current experimental parameters, still all platforms except the Calcium ions enter the repeater regime. In this case, only the scheme based on Rubidium atoms shows the full repeater advantage with a rate scaling $\sim\sqrt{\eta}$ over 400 km.





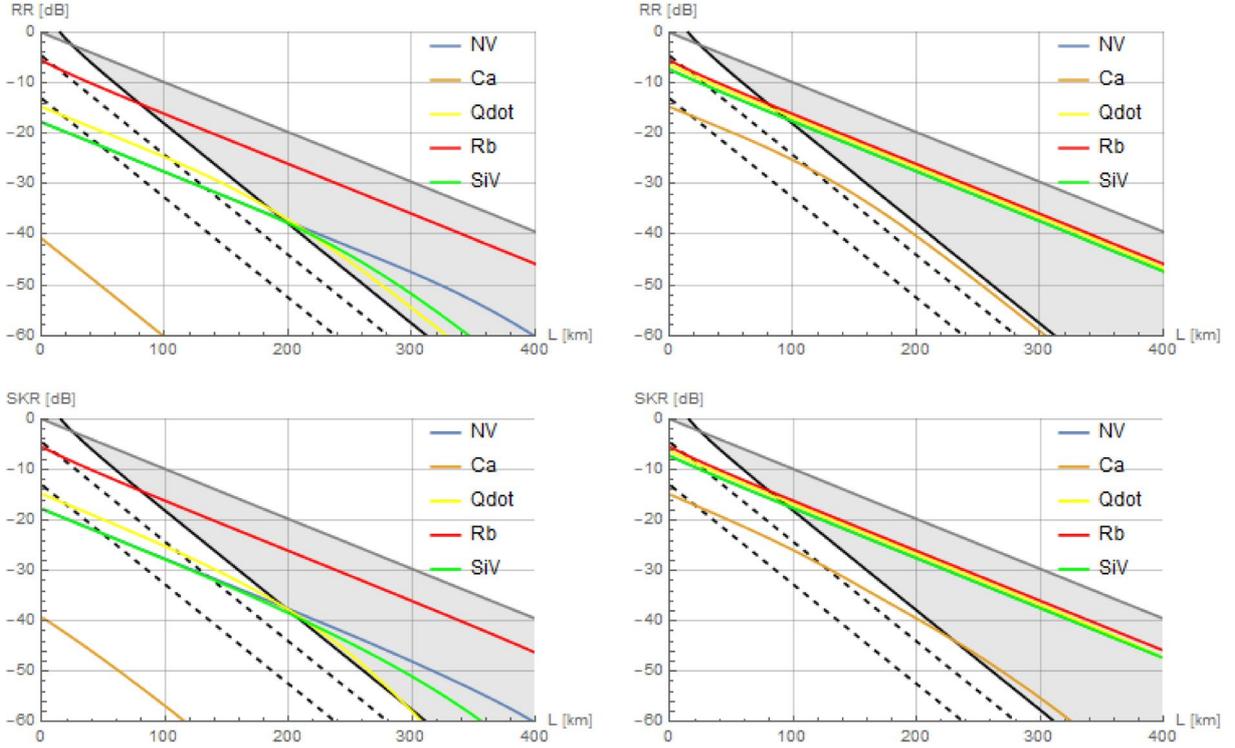

**Fig. 7:** *Secret Key Rates (SKR) and High-Fidelity Raw Rates (RR) for small NRP-based QR schemes (QR cell assuming $P_{write} = 1$ in $P_{link} = P_{source} P_{write}$ ). The bottom plots show SKR in dB as a function of the total distance L in km for experimental parameters as currently available (left) and as potentially available in the future (right). The top plots show RR in schemes where the entangled states effectively created over the total distance L have a fidelity of at least 0.95 (left: current parameters, right: future parameters). The different platforms correspond to NV (violet) and SiV (green) centers, Calcium ions (brown), Rubidium atoms (red), and quantum dots (yellow). The NV curve is invisible for future parameters, but coincides with that of the SiV platform. The light grey area illustrates the (secret key) rate regime between $\sim\eta$ (curve in bold black: "repeaterless" bound) and $\sqrt{\eta}$ (line in dark grey: optimal rate for QR cells or two-segment QR schemes). The bold black dashed lines represent the realistic "repeaterless" bound $P_{link}\eta/2$ (direct transmission via PPL) with finite link efficiencies $P_{link} = 0.1, 0.7$.*

For the NRP-QR cell we may also consider an explicit write-in mechanism in the form of a linear optical BM (Fig. 6b). In this case, instead of assuming unit write-in efficiency like for the rates calculated in Fig. 7, we have $P_{write} = \frac{1}{2} P_{source} (\eta_{det})^2$ as mentioned above. We present the corresponding rates calculated for this situation in Sec. 5 of the Supporting Information.





**4. Conclusion**

As the effective clock rate in a memory-based QKD or QR system is always slower than that of a direct point-to-point quantum connection driven from a laser source at ~GHz rates, the memory-based system will become potentially more efficient only at large communication distances requiring sufficiently many elementary QR segments and additional quantum error detection and correction at higher "nesting levels" of the QR. At such large scales, quantum memories must be sufficiently long-lived or fault-tolerant to survive the necessary waiting times especially for the classical signals sent back and forth between the QR stations. However, a necessary requirement for a large-scale QR to show a performance superior to that of direct transmission is that its fundamental elements already exceed the bounds constraining a "repeaterless" system on a smaller scale: employing an elementary QR cell or a two-segment QR should on average lead to a larger secret key or qubit transmission rate than obtainable in a direct transmission. We have investigated such basic elements for a QR system considering two protocol variants for three different hardware platforms.

Combining the basic building blocks in a modular fashion allows to construct a QR system that is, considering only channel loss, scalable to larger distances. For the realistic situation including memory and depolarizing errors (e.g. for an imperfect spin-spin BM) eventually additional methods of quantum error correction/detection will be required. Nonetheless, for the small-scale QR elements (cells and two-segment schemes) discussed in this work the impact of both finite link and memory efficiencies (the latter described by a simple dephasing model including a "memory cutoff") on the repeater performance has been analyzed for various hardware platforms. The aim was to keep our model sufficiently simple in order to allow for an analytic treatment and to be able to assess the performances in terms of a small set of experimental parameters. While, depending on the protocol, some platforms turn out to





be superior to others with current and future experimental parameters as assumed in our model, a promising further direction could be a hybridization between the different platforms, for instance, combining the high clock rates of quantum-dot-based sources with the long memory coherence times of Rubidium atoms or NV centers. In our NRP protocol, where quantum memories can receive photons at a rate only limited by the source's clock rate and the memory write-in and reset times, but not by the classical communication times, the "repeaterless" bounds can be exceeded quite comfortably under the assumptions of our simplified model. Even when NRP-based QR cells are connected to reach larger distances, like in our NRP-based two-segment QR scheme using sources of entangled photon pairs, high source clock rates can still be of great benefit.[48] Yet, in general, once QR building blocks are connected to construct a larger system composed of many repeater segments or cells, the classical communication times become a limiting factor in any protocol based on quantum memories.

Ultimately, deciding which quantum communication system performs better for a given range must rely upon rates determined in Hz, i.e. per time in seconds. Nonetheless, for a sufficiently large range, the better effective transmission efficiency of a memory-based QR system that becomes manifest in a scaling-with-distance advantage over any point-to-point link will eventually also lead to higher rates in Hz for the QR. In particular, combining many sufficiently short repeater segments improves the scaling and allows to keep the classical communication times small, provided that errors beyond transmission loss can be dealt with via additional quantum error correction. The resulting rates may still be rather small for a single repeater chain, but they can be increased by operating many chains in parallel or via more advanced multiplexing techniques. Such approaches, besides quantum error correction, can also help to keep memory errors small, thus enhancing the overall secret key rates.





**Supporting Information**

Supporting Information is available from the Wiley Online Library or from the author.



Acknowledgements:

The authors acknowledge support from the BMBF in Germany for the project Q.Link.X.

Received: ((will be filled in by the editorial staff))
Revised: ((will be filled in by the editorial staff))
Published online: ((will be filled in by the editorial staff))

# Supporting Information

**Extending Quantum Links: Modules for Fiber- and Memory-Based Quantum Repeaters**

*Peter van Loock\*, Wolfgang Alt, Christoph Becher, Oliver Benson, Holger Boche, Christian Deppe, Jürgen Eschner, Sven Höfling, Dieter Meschede\*, Peter Michler, Frank Schmidt, Harald Weinfurter*

## S1. Graphical language, experimental parameters, and figures of merit

Here we summarize the graphical symbols as used in this paper, which we propose for a visual representation of the structure and the protocols of QR links.

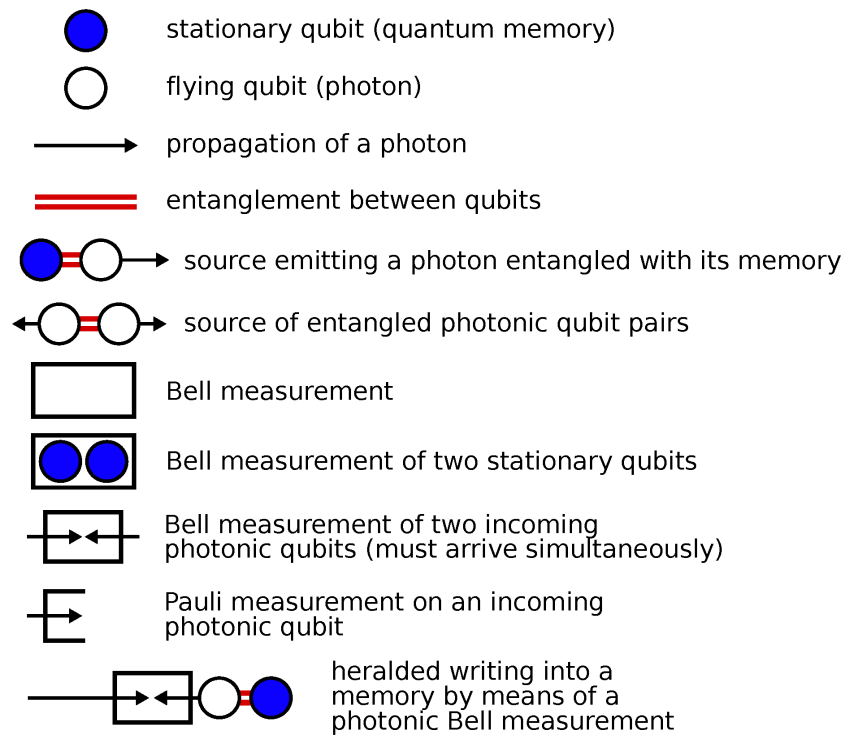





We further summarize the most important experimental parameters and the figures of merit to assess the performance of a QR link.

$P_{link}$       zero-length coupling efficiency, link coupling efficiency

$\tau_{clock}$       source/memory clock time (inverse clock rate)

$\tau_{coh}$       memory coherence time

$\eta$       fiber channel transmission efficiency,
          amplitude damping parameter for a single-mode loss channel

$\mathcal{R}$       raw rate in Hz (number of qubits transmitted per time and per mode)

$R$       raw rate (number of qubits transmitted per channel use),
          inverse average number of qubit transmission attempts needed for one success

$R_{link}$       multi-mode link efficiency,
          raw rate (number of qubits transmitted in link per channel use)

$T_0$       time duration for one channel use,
          time duration for one transmission/distribution attempt

SKR       secret key rate (number of secret bits per channel use and per mode)

RR       raw rate with fidelity bound (number of qubits/ebits per channel use and per mode)

## S2. Memory dephasing model including cutoff and secret key rates for QKD

The memory error model we shall consider is pure memory dephasing as described by

$$\rho \to \frac{1}{2}\left(1 + \exp\left(-\frac{t}{\tau_{coh}}\right)\right)\rho + \frac{1}{2}\left(1 - \exp\left(-\frac{t}{\tau_{coh}}\right)\right)Z\rho Z,$$

where $\frac{1}{2}\left(1 - \exp\left(-\frac{t}{\tau_{coh}}\right)\right)$ is the probability for a Pauli-$Z$ phase-flip to occur on the state of a single memory qubit.

For the case of two QR segments or, equivalently, a QR cell with two half segments, we define a random variable $M$ as $|X_1 - X_2|$ where $X_1$ and $X_2$ are independent geometrically distributed random variables describing the number of attempts until success in a single (half)





segment. This means the random variable $M$ counts the number of time steps for which either one of the two memories (i.e. the first memory whose link has been successfully established via detection of a transmitted photon) has to wait for the other one that still attempts to be connected. The waiting quantum memory is subject to dephasing for a duration of $MT_0$. Here $T_0$ is the time duration per attempt whose value is protocol-dependent and, for simplicity, two additional protocol-dependent extra units of dephasing, $2T_0$, are omitted in $M$ (in the quantitative rate analysis and in the plots for the NSP protocol, these two units are included, see below).

Either of the protocols as described in the main text can be effectively treated like an entanglement swapping (quantum teleportation) process in which a final effective entangled state emerges after the BM on the two quantum memories at the central node. Considering a suitable Pauli correction (depending on the BM result) and tracing out the two measured memories, this final state takes the form of

$$\frac{1}{2}\left(1 + \exp\left(-M\frac{T_0}{\tau_{coh}}\right)\right)|\phi^+\rangle\langle\phi^+| + \frac{1}{2}\left(1 - \exp\left(-M\frac{T_0}{\tau_{coh}}\right)\right)|\phi^-\rangle\langle\phi^-|,$$

where $|\phi^\pm\rangle$ are the two two-qubit Bell states $|\phi^\pm\rangle = (|00\rangle \pm |11\rangle)/\sqrt{2}$.

We remark that depending on the protocol and the application we may not actually prepare such an entangled state (for instance, physically present in two spatially separated quantum memories). Instead, in the QKD context, we convert e.g. the usual BB84 protocol that does not rely on physically distributing entangled states into an equivalent entanglement-based QKD protocol, thus simplifying the theoretical analysis. This equivalence can be understood in the following way. Suppose Alice prepares the state $|\phi^+\rangle$ and sends one half to Bob. After its arrival, Alice and Bob perform $X$- and $Z$-measurements on their halves of the entangled state. Then Alice's measurement acts only on the Hilbert space of her qubit and therefore it





commutes with Bob's measurement and possible attacks by Eve. Consequently, she could also perform her measurement before she sends her half to Bob, which is equivalent to preparing and sending BB84 states to Bob. Also notice that the BM on the memories takes place after two successful detections and therefore the Pauli correction can be applied simply on the level of the classical post-processing of the measurement data. We need to save all measurement results and any information about the state preparations and in the end we can discard the information for those cases where the transmission failed.

For the probability distribution of the random variable $M$ we obtain (here $p$ is the success and $q=1-p$ the failure probability for one attempt)

$$\mathbb{P}(M = 0) = \sum_{k=1}^{\infty} \mathbb{P}(X_1 = X_2 = k) = \sum_{k=1}^{\infty} p^2 q^{2(k-1)} = \frac{p}{2-p},$$

and for $j > 0$,

$$\mathbb{P}(M = j) = \sum_{k=1}^{\infty} 2p^2 q^{2(k-1)+j} = \frac{2pq^j}{2-p},$$

where the factor 2 comes from the fact that both cases $X_1 > X_2$ and $X_2 > X_1$ are possible. This allows us to calculate the following expectation value,

$$\mathbb{E}\left(\exp\left(-M \frac{T_0}{\tau_{coh}}\right)\right) = \frac{p}{2-p}\left(\frac{2}{1 - q\exp\left(-\frac{T_0}{\tau_{coh}}\right)} - 1\right),$$

and by summing only up to a cutoff constant $m$ instead of infinity, including a renormalization of the probability distribution, one can easily obtain the expectation value for protocols which abort after the memory has dephased for a predetermined, given number of time steps (attempts). Again note that, depending on the protocol, the overall state may be subject to dephasing for an additional constant amount of $2T_0$. In the case of the NSP protocol, we first generate entanglement between the memory and a photon, and as the next step we send this





photon to a detector over a distance $L_0 = L/2$. Then the detector sends a classical signal to the memory announcing whether the photon was detected or not. Therefore, we have to wait for a time unit of $T_0 = 2L_0/c = L/c$ until we can decide which action should be applied to the memory: storage of the qubit or initialization for a new attempt. Hence, the memory would always decohere for at least one such time step, even in the case when the very first attempt is already successful. Since this argument applies to both memories, the total state decoheres (is subject to dephasing) for $M + 2$ time steps, each with duration $T_0 = L/c$.

However, if we consider the NRP protocol, we send photons to the memory and therefore the memories (almost) immediately know when a transmission was successful. As a consequence, there is no additional constant dephasing in this case and $T_0$ is simply given via the repetition rate of the photon source or the local processing times including the write-in time, whichever is longer.

Using the BB84 protocol,[12] we obtain an ideal asymptotic secret key fraction of $1 - h(e_x) - h(e_z)$, where $h(x) = -x \log_2(x) - (1 - x) \log_2(1 - x)$ is the binary entropy and $e_x, e_z$ are the error rates in the $X$ and $Z$ basis, respectively. Since the $Z$-error rate is equivalently given by the probability to obtain the effective state $|\psi^\pm\rangle$, one can easily see that $e_z$ is zero in our error model. Similarly, the $X$-error rate is given by the probability to obtain $|\psi^-\rangle$ or $|\phi^-\rangle$ and is therefore given by $\frac{1}{2}\left(1 - \mathbb{E}(\exp\left(-M\frac{T_0}{\tau_{coh}}\right))\right)$ up to the protocol-dependent constant dephasing. Hence the asymptotic secret key fraction is given by

$1 - h\left(\frac{1}{2}\left(1 - \mathbb{E}(\exp\left(-M\frac{T_0}{\tau_{coh}}\right))\right)\right)$, and the final secret key rate is then the product of the raw rate (the so-called "yield") and this secret key fraction.

Also notice that the binary entropy function takes on its maximum of 1 when the argument of the function is $\frac{1}{2}$. Thus, we always obtain a non-zero secret key fraction, which is a specific feature of our error model. If we also consider additional error sources like, for example, imperfect (though still deterministic) BMs on the memories, we typically have non-zero error rates in both the $X$ and the $Z$ basis (unlike the sole phase-flip error in the effective entangled state above). Therefore, the secret key fraction can become zero and we typically get more demanding requirements for the memory coherence times.

---

[12] We consider the biased BB84 scheme here where one of the two bases is employed more often than the other which, in the asymptotic limit of infinite repetitions, allows to remove the ½ factor in the rates of standard BB84 and increase the sifting factor to unity [49].

## S3. Calculation of raw rates

The performance of a QR may be quantified by the secret key rate that can be obtained for a given length $L$ of the quantum channel connecting the two parties Alice and Bob who aim to securely communicate with each other. Besides the secret key fraction, for calculating the (asymptotic) secret key rate, we need an expression for the raw rate, i.e. in our case, the number of quantum bits that can be transmitted over a lossy channel of length $L$, employing that channel once and sending one optical mode through that channel (i.e. "per channel use" and "per mode"). As the memory-based QR has at least one intermediate station as opposed to a PPL for direct transmission, it may not be immediately obvious how to count the channel uses. In our case, one channel use corresponds to one attempt to establish a link, and because the two (half) segments can be simultaneously attempted to be bridged, the total number of attempts, on average, to transmit one qubit over the entire distance can be expressed by $\mathbb{E}(\max(X_1, X_2))$. The probability for successfully transmitting one qubit can then be written as $1/\mathbb{E}(\max(X_1, X_2))$. This then corresponds to the number of qubits transmitted per channel use, i.e. a dimensionless raw rate expressed per channel use.

The effect of imperfect quantum memories, i.e., quantum memories with finite coherence times (see the dephasing model of the preceding section), can be taken into account in the raw rate by imposing a maximally allowed storage time of the loaded memory at the central station waiting for a second transmission to succeed. In other words, the QR protocol is aborted as soon as a quantum memory's storage time limit is exceeded. If this "cutoff" is chosen to be well below the memory's coherence time, one can ensure that the quality of the entangled light-matter state is still so high and hence that of the final (effective) entangled





state too, such that errors are negligible. In the QKD context, this corresponds to a secret key fraction near unity. However, such an approach would be at the expense of the raw rate, because aborting and restarting the protocol more frequently for a small cutoff time means that it takes longer to finally distribute a qubit over the total distance, thus reducing the raw rate. Due to this trade-off, there is an optimal cutoff that maximizes the secret key rate. Nonetheless, we shall also consider sufficiently small cutoffs that lead to fidelities of the final (effective) entangled states that are above a certain fidelity value. This may also be relevant for applications different from QKD. Generally, smaller memory coherence times and thus shorter storage time limits require a correspondingly faster abortion and restart of the protocol leading to a smaller transmission probability. For the NSP protocol, this effect depends on the total distance $L$, because for larger $L$, the required storage time per transmission attempt grows such that for a given, fixed memory coherence time the effective memory efficiency drops, which becomes visible in the QR performance. As a consequence, in this case, the cutoff becomes distance-dependent in order to keep the fidelity above a certain threshold and the maximal secret key rates have smaller optimal cutoffs for larger distances. In the NRP protocol, this $L$-dependence disappears, because the quantum signals are sent to, and no longer emitted from, the quantum memories, in which case the duration of every transmission attempt only depends on the source's repetition rate and the local processing / write-in times, and no longer on the distance between memories and detectors.

Calculating the expression $1/\mathbb{E}(\max(X_1, X_2))$, the dimensionless raw rate (or qubit transmission probability) for a memory-based scheme with one central memory node including memory cutoff time is given by [22]

$$R(m) = \frac{p \left[2 - p - 2q^{m+1}\right]}{3 - 2p - 2q^{m+1}} \; P_{\text{BM}} \; .$$

Here, $p$ and $q$ are again the success and failure probabilities of a single attempt in one (half) segment of length $L/2$. Thus, for deterministic local state preparations (or, more generally,





unit link coupling efficiencies), we have $p = \sqrt{\eta}$. The final BM efficiency on the two memories is included via the extra factor $P_{BM}$, which can be set to one for a deterministic BM ($P_{BM} = 1$ in the following). The parameter $m$ determines the maximal acceptable number of attempts (the above-mentioned memory cutoff) a loaded memory is allowed to wait for a second successful transmission attempt. Note that for $m = 0$ we obtain the no-memory case, corresponding to $R(0) = p^2 = \eta$, which is just the result one obtains for direct transmission, i.e. the "repeaterless" bound for distance $L$ (for not too small $L$). Conversely, for $m \to \infty$ (corresponding to the perfect memory case with no need for aborting the protocol), we have $R \to \frac{p(2-p)}{3-2p} \equiv R(m \to \infty)$, which, for small $p$ becomes approximately $R \approx \frac{2}{3} p \sim \sqrt{\eta}$ (and this scaling becomes $\eta^{1/n}$ for $n$ repeater segments). The $\sqrt{\eta}$-scaling corresponds to the optimal transmission in a memory-based QR with a single node or, equivalently, two segments.

## S4. Additional results: two-segment QR in the NSP protocol

In comparison to the rates of the NSP-QR cell (illustrated by Fig. 4b) as shown in Fig. 5, below we also present the rates calculated for the two-segment QR as illustrated by Fig. 4a. The subtle differences between these two small-scale QR variants are discussed in the main text. In addition to the short discussion there, let us emphasize here that for a reasonable comparison, we did not include dephasing errors on the outer memories (those most left and right in Fig. 4a). Practically, in the context of QKD, this means that Alice and Bob would immediately measure their qubits and not store any quantum states at all; thus, storage again takes place only at the central node. On the other hand, such an approach prevents the two-segment scheme from its possible use beyond QKD, because the two-segment scheme is potentially more versatile compared with the NSP-QR cell when the outer memories of the two segments are also exploited for quantum storage.





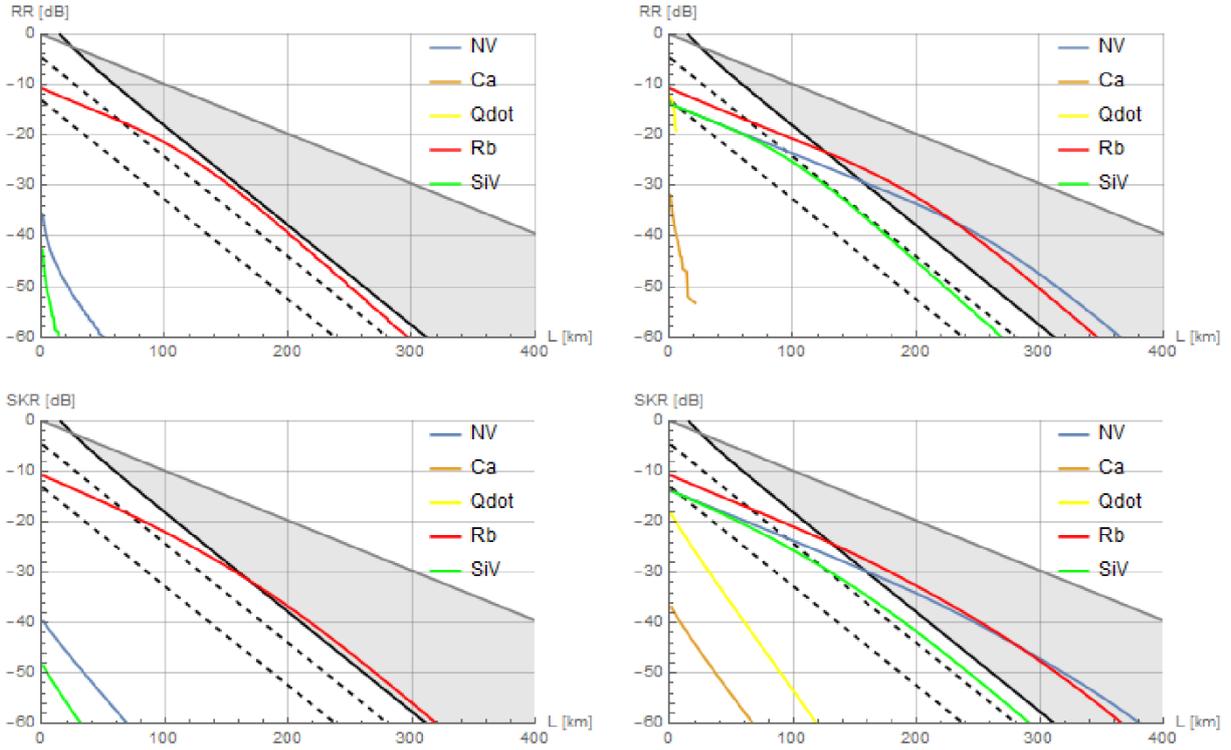

**Fig. S1:** *Secret Key Rates (SKR) and High-Fidelity Raw Rates (RR) for a small NSP-based QR scheme (two-segment QR). The bottom plots show SKR in dB as a function of the total distance L in km for experimental parameters as currently available (left) and as potentially available in the future (right). The top plots show RR in schemes where the entangled states effectively created over the total distance L have a fidelity of at least 0.95 (left: current parameters, right: future parameters). Curves that are disappearing beyond certain distances (or completely missing) no longer (never) exceed F=0.95. The different platforms correspond to NV (violet) and SiV (green) centers, Calcium ions (brown), Rubidium atoms (red), and quantum dots (yellow). The light grey area illustrates the (secret key) rate regime between $\sim\eta$ (curve in bold black: "repeaterless" bound) and $\sqrt{\eta}$ (line in dark grey: optimal rate for QR cells or two-segment QR schemes). The bold black dashed lines represent the realistic "repeater-less" bound $P_{link}\eta/2$ (direct transmission via PPL) with finite link efficiencies $P_{link} = 0.1$, 0.7.*

One can see that overall the curves are very similar for the two QR variants with a visibly better performance of the QR cell. For the two-segment scheme as shown above, the secret key rate of the Rubidium-atoms-based platform now only barely enters the repeater regime with current parameters and for future parameters the NV-center-based and Rubidium-based rates, though clearly entering the repeater regime for the simple model considered, are a bit worse compared to their corresponding rates with the QR cell. The SiV platform no longer reaches the repeater regime, not even for future parameters as it did before with the QR cell.





## S5. Additional results: Bell-state measurement-assisted memory write-in (NRP)

In comparison to the rates of the NRP-QR cell with ideal unit write-in efficiency as shown in Fig. 7, below we also present the rates calculated for a scheme with quantum teleportations of the arriving photonic qubits onto the spin qubits with the help of locally prepared spin-photon entangled states and linear optical BMs (see Fig. 6b). With future parameters, except for the Calcium ions, the repeater regime can still be entered and the repeater rate slopes well maintained over 400 km despite the non-unit write-in efficiency.

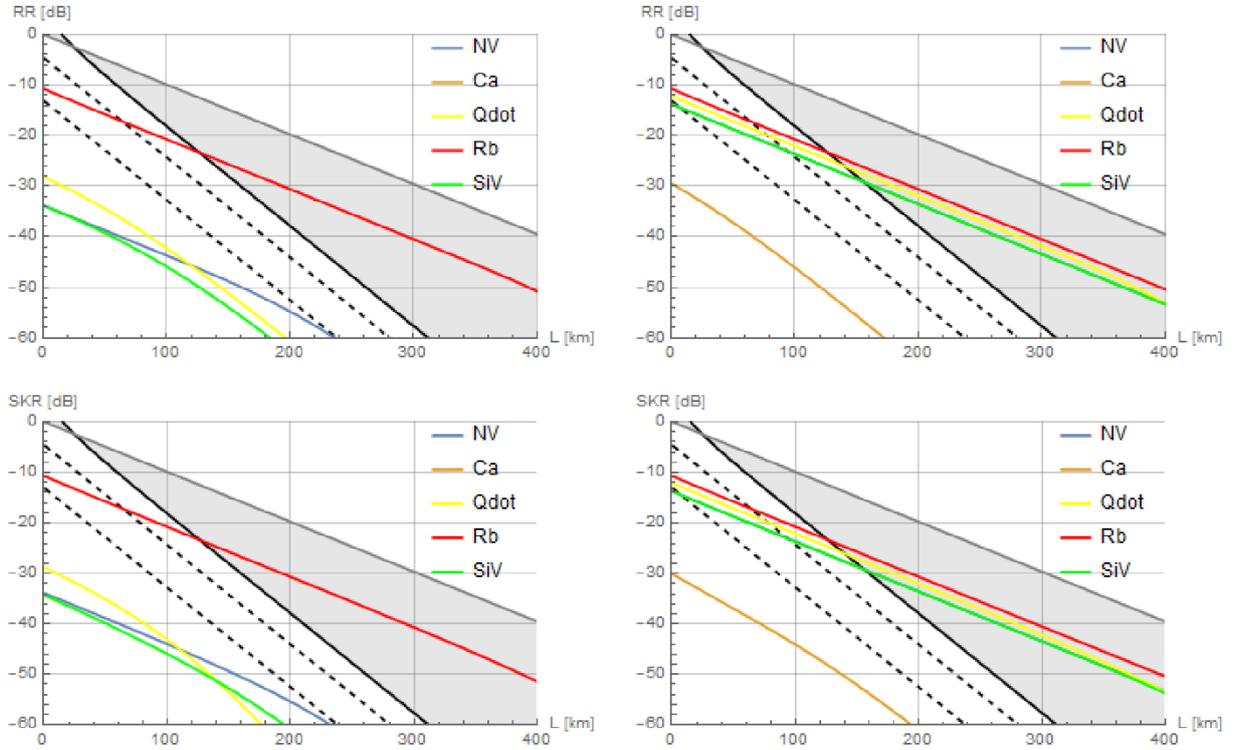

**Fig. S2:** *Secret Key Rates (SKR) and High-Fidelity Raw Rates (RR) for a small NRP-based QR scheme (QR cell with linear optical teleportation-assisted memory write-in). The bottom plots show SKR in dB as a function of the total distance L in km for experimental parameters as currently available (left) and as potentially available in the future (right). The top plots show RR in schemes where the entangled states effectively created over the total distance L have a fidelity of at least 0.95 (left: current parameters, right: future parameters). Curves that are completely missing for Calcium atoms never exceed F=0.95. The different platforms correspond to NV (violet) and SiV (green) centers, Calcium ions (brown), Rubidium atoms (red), and quantum dots (yellow). The NV curve is invisible for future parameters, but coincides with that of the SiV platform. The light grey area illustrates the (secret key) rate regime between ~η (curve in bold black: "repeaterless" bound) and √η (line in dark grey: optimal rate for QR cells or two-segment QR schemes). The bold black dashed lines represent the realistic "repeater-less" bound $P_{link}\eta/2$ (for direct transmission via PPL) with finite link efficiencies $P_{link} = 0.1, 0.7$.*